\documentclass[twocolumn,showpacs,aps,amssymb]{revtex4}

\usepackage{graphicx}
\usepackage{dcolumn}
\usepackage{bm}

\begin{document}

\title{Oscillation damping of chiral string loops}

\author{Eugeny Babichev}
 \email{babichev@inr.npd.ac.ru}
\author{Vyacheslav Dokuchaev}
 \email{dokuchaev@inr.npd.ac.ru}
 \affiliation{Institute For Nuclear Research of the Russian Academy of
Sciences \\
 60th October Anniversary Prospect 7a, 117312 Moscow, Russia}

\date{\today}

\begin{abstract}

Chiral cosmic string loop tends to the stationary (vorton) configuration
due to the energy loss into the gravitational and electromagnetic
radiation. We describe the asymptotic behaviour of near stationary chiral
loops and their fading to vortons. General limits on the gravitational
and electromagnetic energy losses by near stationary chiral loops are
found. For these loops we estimate the oscillation damping time. We
present solvable examples of gravitational radiation energy loss by some
chiral loop configurations. The analytical dependence of string energy
with time is found in the case of the chiral ring with small amplitude
radial oscillations.
\end{abstract}



\maketitle

\section{Introduction}
\label{sec:intro}

 Cosmic strings are one-dimensional topological defects
which may have formed during phase transition in the early universe
\cite{Kibble1, Zel'dovich1, Vilenkin1}. Witten \cite{Witten1} has shown
that strings could be superconducting in certain particle physics models.
The presence of current in a string leads to the principal specific
feature: the superconducting string loop may form a stable stationary
configuration \cite{Davis1, Haws1, Haws2}. Cosmic strings lose their
energy on gravitational and electromagnetic radiation (if string is
superconducting). As a result, the ``ordinary'' not extremely long cosmic
strings without the current evaporate completely during the cosmological
time. On the contrary the superconducting string loops could survive due
to the presence of conserved ``charge'' and tend to the stable
configuration which is named the chiral vorton \cite{Davis1}.

Numerous works were devoted to calculations of the gravitational and
electromagnetic radiation by cosmic strings (see reviews and references
in \cite{Vilenkin2, Kibble2}). Unfortunately the general problem of the
motion of a superconducting cosmic string coupled to the electromagnetic
field is not solved analytically. The using of Nambu-Goto equations of
string motion in general case results in the singular cusp formation and
the divergence of the electromagnetic power radiated by the string.
Nevertheless equations of motion can be solved precisely \cite{Carter1,
Davis2, Vilenkin3} if (i) the gauge field influence on the string motion
is negligible, e.g. when the superconducting current is neutral, and (ii)
the string current $J^{a}$ is chiral, i.~e. $J^{a} J_{a} = 0$, where $
J^{a}$ is a two-dimensional current on the string world surface. It turns
out that electromagnetic power radiated by the cusp of chiral cosmic
strings is finite \cite{Blanco}.

Self-consistent calculations of string evolution with time is complicated
by the influence of the radiation back reaction on the string motion (see
some approaches to numerical calculations of back reaction in
\cite{Quashnock}). In this work we consider the electromagnetic and
gravitational radiation by chiral strings loops which are close to the
stationary vorton state (when amplitude of loop oscillations is very
small). In this case it is physically justified the supposition that all
string oscillations are faded out. We determine the upper bounds on
gravitational and electromagnetic radiation of near stationary chiral
loops when oscillations are small in amplitude. For some simple
configurations of the string loops we calculate both types of radiation.
We also find the analytical law for behavior in time of the string loop
energy and current basing on the symmetries of the problem in the case
when the final vorton configuration is the chiral ring and oscillations
are radial. For some other less symmetric examples of loops we estimate
the damping time of small amplitude oscillations. It turns out that in
all considered examples the oscillation damping time of chiral strings
due to gravitational radiation is order of magnitude longer than the
known lifetime estimations for the ``ordinary'' cosmic strings without
the current.

The paper is organized in the following order. Section~\ref{sec:II}
describes the general properties of the motion of chiral string loop in
flat space-time. In Section~\ref{sec:III} we consider the gravitational
and electromagnetic radiation of the chiral string loops. The radiation
power of the near stationary chiral loop with small amplitude
oscillations is determined and the corresponding upper bounds for
gravitational and electromagnetic radiation are found. Some solvable
examples of near stationary chiral strings including radially oscillating
ring are presented in Section~\ref{sec:IV}. In Section~\ref{sec:V} the
damping time of weakly oscillating chiral loops is estimated and the
exact time evolution law for radially oscillating chiral ring is found.
Section~\ref{sec:VI} briefly summarizes some features of the oscillation
damping of chiral loops.

\section{Chiral string motion}
\label{sec:II}

We remind here the known basic properties of chiral string motion in flat
space-time required for the following calculations. The moving string
sweeps a two-dimensional world-sheet in the Minkowski space-time. The
corresponding space-time points on this world-sheet
$x^{\mu}=x^{\mu}(\sigma^{a})$, where indices $a=0,1$, and $\sigma^{0}$ and
$\sigma^{1}$ are coordinates on a two-dimensional world-sheet. If the
back reaction of electromagnetic and gravitational radiation is
negligible, then equations of motion of the string with a current is given
by \cite{Witten1,Vilenkin4}
\begin{eqnarray}
  \label{in:motion1}
  \partial_{a}\left[\sqrt{-\gamma}\left(\mu \gamma^{ab}+\theta^{ab}
  \right)x^{\nu}_{,b}\right] =0, \nonumber \\
  \partial_{a}\left(\sqrt{-\gamma}\gamma^{ab}\phi_{,b}\right)=0,
\end{eqnarray}
where $\gamma_{ab}=\eta_{ab} x^{\mu}_{,a} x^{\nu}_{,b}$ is the induced
metric on a two-dimensional world-sheet, $\mu$ is the energy of a string
per unit length, $\phi$ is the auxiliary scalar field and $\theta^{ab}$
is the energy-momentum tensor of the charge carriers
\begin{equation}
  \label{in:theta}
  \theta^{ab}=\gamma^{ac}\gamma^{bd}\phi_{,c}\phi_{,d}
  -\frac{1}{2}\gamma^{ab}\gamma^{cd}\phi_{,c}\phi_{,d}\,.
\end{equation}
The conserved current on a two-dimensional world-sheet is written
in terms of the scalar field
\begin{equation}
  \label{in:J1}
  J^{a}=\frac{q}{\sqrt{-\gamma}}\epsilon^{ab}\phi_{,b}\,,
\end{equation}
where $q$ is the  coupling constant of the fields in the string. The
four-dimensional current is expressed through (\ref{in:J1}) as
\cite{Vilenkin4}
\begin{equation}
  \label{in:j1}
  j^{\mu}(x)= \int d^{2}\sigma \sqrt{-\gamma} J^{a}
  x^{\mu}_{,a}\delta^{(4)}(x-x(\sigma)).
\end{equation}
If this current is chiral, i.~e.
\begin{equation}
 \label{in:chirality}
  J^{a}J_{a}=0,
\end{equation}
then equations of motion (\ref{in:motion1}) can be solved analytically
\cite{Carter1, Davis1, Vilenkin3}. This solution for the chiral string
with an invariant length $L$ is
\begin{equation}
  \label{in:motion2}
  x^{0}=t, \quad
  \mathbf{x}(t,\sigma)=
  \frac{L}{4\pi}\left[\mathbf{a}(\xi)+\mathbf{b}(\eta)\right],
\end{equation}
where $\mathbf{a}(\xi)$ and $\mathbf{b}(\eta)$ are the vector functions
of $\xi=(2\pi/L)(\sigma-t)$ and $\eta=(2\pi/L)(\sigma+t)$ obeying the
constraints:
\begin{equation}
  \label{in:conab1}
  {\mathbf{a}'}^{2}=1,\quad
  {\mathbf{b}'}^{2}\leq 1
\end{equation}
For closed strings (with identified ends) the vector functions
$\mathbf{a}(\xi)$ and $\mathbf{b}(\eta)$ define loops which are called
$a$- and $b$- loops. The corresponding solution for scalar field is
\begin{equation}
  \label{in:phi}
  \phi(\sigma,t)=\frac{L}{2\pi} F(\eta),
\end{equation}
where $F(\eta)$ is an arbitrary function. It is useful to define
${\mathbf{b}'}^{2}\equiv k^{2}(\eta)$. Then new function $k(\eta)$ is
expressed through $F(\eta)$ in the following way:
\begin{equation}
  \label{in:k1}
  k^{2}(\eta)=1-\frac{4F'^{2}(\eta)}{\mu}.
\end{equation}
At $k=0$ the chiral string becomes vorton.

The energy-momentum tensor of the string in this gauge
\begin{equation}
  \label{in:T1}
  T^{\mu\nu}= \mu\int
  d\sigma\left(\dot{x}^{\mu}\dot{x}^{\nu}-x'^{\mu}x'^{\nu}\right)
  \delta^{(3)}\left(\mathbf{x}-\mathbf{x}(\sigma,t)\right),
\end{equation}
and correspondingly the string total energy
\begin{equation}
  \label{in:E1}
  E=\mu\int d\sigma,
\end{equation}
with $\sigma$ to be the measure of the string total energy. The
four-dimensional current (\ref{in:j1}) can be written as
\begin{equation}
  \label{in:j2}
  j^{\mu}(\mathbf{x},t)=q \int d\sigma
  \phi'(x'^{\mu}-\dot{x}^{\mu})
  \delta^{(3)}\left(\mathbf{x}-\mathbf{x}(\sigma,t)\right),
\end{equation}
with the total charge for the closed string loop
\begin{equation}
  \label{in:Q0}
  q_{\rm tot}=\frac{q\sqrt{\mu}}{2}\int d\sigma\sqrt{1-k^{2}}.
\end{equation}

\section{Chiral string radiation}
\label{sec:III}

\subsection{Gravitational radiation}

In the case of the periodic physical systems with period $T$ for the
energy-momentum tensor $T_{\mu\nu}$ we will have
$T^{\mu\nu}(\mathbf{x},t)=T^{\mu\nu}(\mathbf{x},t+T)$. Fourier series for
$T_{\mu\nu}$ is
\begin{eqnarray}
  \label{r:serTj1}
  T^{\mu\nu}(\mathbf{x},t)&=&
  \sum_{l=1}^{\infty}e^{-i\omega_{l}t}T^{\mu\nu}(\omega_{l},\mathbf{x})
  +{\rm c.c},
\end{eqnarray}
where \ ${\rm c.c}$ \ denotes the complex conjugation of the preceding
expression, $\omega_{l}=2\pi l/T$ is the frequency and
\begin{eqnarray}
  \label{r:serTj2}
  T^{\mu\nu}(\omega_{l},\mathbf{x})&=&
  \frac{1}{T}\int_{0}^{T}dte^{i\omega_{l}t}T^{\mu\nu}(\mathbf{x},t).
\end{eqnarray}
For $T^{\mu\nu}(\omega_{l},\mathbf{x})$ we have Fourier transform
\begin{eqnarray}
  \label{r:FourierTj}
  \hat{T}^{\mu\nu}(\omega_{l},\mathbf{n})&=&
  \int d^{3}x T^{\mu\nu}(\omega_{l},\mathbf{x})
  e^{-i\omega_{l}\mathbf{n}\mathbf{x}}
\end{eqnarray}
where $\mathbf{n}=\mathbf{x}/|\mathbf{x}|$.

The corresponding gravitational power, radiated per unit solid angle
$d\Omega$ by periodic system is given by \cite{Weinberg}
\begin{equation}
  \label{r:sumE}
  \frac{d \dot{E}^{\rm gr}}{d \Omega}=
  \sum_{l=1}^{\infty}\frac{d\dot{E}^{\rm gr}({\omega}_{l})}{d\Omega},
\end{equation}
where
\begin{equation}
  \label{r:E1}
  \frac{d\dot{E}^{\rm gr}({\omega_{l}})}{d\Omega}
  =\frac{G\omega^{2}_{l}}{\pi}P_{ij}P_{lm}[\hat{T}^{*}_{il}\hat{T}_{jm}
  -\frac{1}{2}\hat{T}^{*}_{ij}\hat{T}_{lm}].
\end{equation}
Here $P_{ij}=\delta_{ij}-n_{i}n_{j}$ is the projection operator to the
plane perpendicular to vector $\mathbf{n}$. It is possible to simplify
the equation (\ref{r:E1}) further if rewrite it in the `corotating' basis
$(\mathbf{e}'_{1},\mathbf{e}'_{2},\mathbf{e}'_{3})=
(\mathbf{n},\mathbf{v},\mathbf{w})$, where $\mathbf{v}$ and $\mathbf{w}$
are the arbitrary unit vectors, perpendicular each other and to vector
$\mathbf{n}$. In this corotating basis (\ref{r:E1}) transforms to
\cite{Durrer}
\begin{equation}
  \label{r:E2}
  \frac{d \dot{E}^{\rm gr} ({\omega_{l}})}{d \Omega}=
  \frac{G {\omega}^{2}_{l}}{\pi}[{\tau}^{*}_{pq}
  {\tau}_{pq}-\frac{1}{2}{\tau}^{*}_{qq}{\tau}_{pp}].
\end{equation}
where  $\tau_{pq}$ are correspondingly the Fourier-transforms of an
energy-momentum tensor in the corotating basis. Only indexes $p,q$ with
values $2$ and $3$ appear in the equation (\ref{r:E2}). The
Fourier-transforms $\tau_{pq}$ can be expressed in the following way
\cite{Durrer}
\begin{equation}
  \label{r:tau1}
  {\tau}_{pq}(\omega_{l},\mathbf{n})=-\frac{L\mu}{2}[I_{p}(l)
  Y_{q}(l)+Y_{p}(l) I_{q}(l)],
\end{equation}
where functions $I_{p}(l)$ and $Y_{q}(l)$ are determined through
''fundamental integrals''
\begin{eqnarray}
  \label{r:IY1a}
  I_{2}(l)&\equiv& I_{l}(\mathbf{n},\mathbf{v}),\quad
  I_{3}(l)\equiv I_{l}(\mathbf{n},\mathbf{w}),\nonumber\\
  Y_{2}(l)&\equiv& Y_{l}(\mathbf{n},\mathbf{v}), \quad
  Y_{3}(l)\equiv Y_{l}(\mathbf{n},\mathbf{w}).
\end{eqnarray}
with
\begin{eqnarray}
  \label{r:I1}
  I_{l}(\mathbf{n},\mathbf{z})&\equiv&
  \frac{1}{2\pi}\int_{0}^{2\pi} d\xi e^{-il(\xi
  +\mathbf{n}\mathbf{a}(\xi))} \mathbf{a}'(\xi)\mathbf{z},
\end{eqnarray}
\begin{eqnarray}
  \label{r:Y1}
  Y_{l}(\mathbf{n},\mathbf{z})&\equiv&
  \frac{1}{2\pi}\int_{0}^{2\pi} d\eta e^{il(\eta
  -\mathbf{n}\mathbf{b}(\eta))} \mathbf{b}'(\eta)\mathbf{z}.
\end{eqnarray}
Using functions (\ref{r:I1}) and (\ref{r:Y1}) we rewrite the gravitational
power (\ref{r:E2}) radiated per unit solid angle at frequency
$\omega_{l}=4\pi l/L$ in the final form
\begin{eqnarray}
  \label{r:E3}
  \frac{d \dot{E}^{\rm gr}_{l}}{d\Omega}&=& 16 \pi l^{2} G\mu^{2}\left\{
  \left|I_{2}(l)Y_{2}(l)\right|^{2}+\left|I_{3}(l)Y_{3}(l)\right|^{2}\right.
  \nonumber \\ & &
  +\frac{1}{2}\left|I_{2}(l)Y_{3}(l)+I_{3}(l)Y_{2}(l)\right|^{2}
  \\ & &
  -\left.\frac{1}{2}\left|I_{2}(l)Y_{2}(l)+I_{3}(l)Y_{3}(l)\right|^{2}
\right\}. \nonumber
\end{eqnarray}

\subsection{Electromagnetic radiation}

We will consider the electromagnetic radiation of chiral strings close to
the stationary configuration, i.~e. when the current is near to its
maximum value. For the opposite limiting case of electromagnetic
radiating by the strings with a weak electromagnetic current see
\cite{Blanco,Berezinsky}.

Similar to the case of the gravitational radiation, the electromagnetic
power emitted per solid angle $d\Omega$ is expanded in Fourier series:
\begin{equation}
  \label{r:sumEe}
  \frac{d \dot{E}^{\rm em}}{d\Omega}=
  \sum_{l=1}^{\infty}\frac{d\dot{E}^{\rm em}({\omega}_{l})}{d\Omega}.
\end{equation}
Expanding the electromagnetic current $j^{\mu}$ in Fourier series we find
\begin{equation}
j^{\mu}(\mathbf{x},t)=
  \sum_{l=1}^{\infty}e^{-i\omega_{l}t}j^{\mu}(\omega_{l},\mathbf{x})
  +{\rm c.c},
\end{equation}
where
\begin{equation}
j^{\mu}(\omega_{l},\mathbf{x})=
  \frac{1}{T}\int_{0}^{T}dte^{i\omega_{l}t}T^{\mu}(\mathbf{x},t).
\end{equation}
Fourier transforms of $j^{\mu}$ are
\begin{equation}
  \hat{j}^{\mu}(\omega_{l},\mathbf{n})= \int d^{3}x
  j^{\mu}(\omega_{l},\mathbf{x})
  e^{-i\omega_{l}\mathbf{n}\mathbf{x}}.
\end{equation}
Then expanding in standard way the retarded solution of Maxwell equations
for electromagnetic potential in Lorenz gauge $\square A_{\mu}=-4\pi
j_{\mu}$ in series of $1/r$, taking the leading term and substituting in
formula for electromagnetic radiation \cite{Landau} $d\dot{E}^{\rm
em}/d\Omega = \mathbf{E}\times\mathbf{H}/4\pi$ one can find finally
\begin{equation}
  \label{r:Ee0}
  \frac{d\dot{E}^{\rm em}(\omega)}{d\Omega}=
  \frac{\omega^{2}}{2\pi}{\iota}^{*}_{p}{\iota}_{p}.
\end{equation}
Here $\iota_{p}$ is the Fourier transform of electromagnetic current in
the corotating basis. In the case of chiral loop $\iota_{p}$ can be
expressed in the following way:
\begin{equation}
  \label{r:iota1}
  {\iota}_{p}(\omega_{l},\mathbf{n})=\frac{Lq\sqrt{\mu}}{2} I_{p}(l) M(l),
\end{equation}
where $I_{p}$ is from (\ref{r:IY1a}) and
\begin{eqnarray}
  \label{r:M}
  M(l)&\equiv& \frac{1}{2\pi}\int_{0}^{2\pi} d\eta e^{il(\eta
  -\mathbf{n}\mathbf{b}(\eta))} \sqrt{1-|\mathbf{b}'(\eta)|^{2}}.
\end{eqnarray}
Using functions (\ref{r:I1}) and (\ref{r:M}) we rewrite the
electromagnetic power, radiated per unit solid angle at frequency
$\omega_{l}$ in the final form
\begin{equation}
  \label{r:Ee2}
  \frac{d\dot{E}^{\rm em}_{l}}{d\Omega}= 2\pi l^{2} q^{2}\mu
  \left(\left|I_{2}(l)\right|^{2}+\left|I_{3}(l)\right|^{2}\right)
  \left|M(l)\right|^{2}
\end{equation}

\subsection{Radiation of near stationary loops}

Now we may consider in more details the small amplitude oscillations of
the chiral string loop, i.~e. when string is near to its stationary
state. This means that arbitrary function $b(\eta)$ in the solution for
string motion (\ref{in:motion2}) is such $b'(\eta)=k(\eta)\ll 1$. If
three-dimensional coordinates are chosen so that a $b$-loop is near to
the origin of coordinate system (e.~g., $b$-loop intersects exactly the
origin of coordinate system) then $b(\eta)\ll 1$. In this case we can
neglect the term $\mathbf{n}\mathbf{b}$ in the first order of expansion
on parameter $k$ in the exponent of (\ref{r:Y1}). As a result we obtain
\begin{eqnarray}
  \label{s:Y1}
  Y_{l}(\mathbf{n},\mathbf{z})\simeq
  \frac{1}{2\pi}\int_{0}^{2\pi} d\eta e^{il\eta}
  \mathbf{b}'(\eta)\mathbf{z}
\end{eqnarray}
and therefore $Y_{l}(\mathbf{n},\mathbf{z})$ in this case is $l$-th
Fourier component $\mathbf{b}'(\eta)\mathbf{z}$. To find the first order
expansion on $k$ in (\ref{r:M}) we should take into account not only
zero, but also the first term in expansion of the exponent
$\exp(-il\mathbf{n}\mathbf{b})$. This gives
\begin{eqnarray}
  \label{s:M}
  M(l)\simeq -i\frac{l}{2\pi}\int_{0}^{2\pi} d\eta
  e^{il\eta}\mathbf{n}\mathbf{b}.
\end{eqnarray}
This provides the opportunity to determine an upper limit of the
gravitational radiation power in the case of near stationary chiral
loops. From (\ref{r:E3}) it follows
\begin{eqnarray}
  \label{s:E1}
  \left|\frac{d\dot{E}^{\rm gr}(\omega_{l})}{d\Omega}\right| \!\!
  &\leq&
  \!\! 8\pi G\mu^{2}l^{2}
  \! \left\{\left[\,
  |I_{2}(l)Y_{2}(l)|\!+\!|I_{3}(l)Y_{3}(l)|\,\right]^{2} \right.
  \nonumber \\
  &+&\!\! \left.
  \left[\,|I_{2}(l)Y_{3}(l)|+|I_{3}(l)Y_{2}(l)|\,
  \right]^{2}\right\}.
\end{eqnarray}
Let us assume now, that $b(\eta)$ is twice continuously differentiable,
and $b'''(\eta)$ is piece-wise continuous. Successive double integration
by parts in (\ref{s:Y1}) gives
\begin{equation}
  \label{s:Y2}
  Y_{l}(\mathbf{n},\mathbf{z})=-\frac{1}{2\pi l^{2}}
  \int_{0}^{2\pi} \! d\eta e^{il\eta} \mathbf{b}'''(\eta)\mathbf{z}.
\end{equation}
From (\ref{s:Y2}) it follows the inequality
\begin{equation}
  \label{s:Y3}
  \left|Y_{l}(\mathbf{n},\mathbf{z})\right|\leq\frac{b_{3}}{l^{2}},
\end{equation}
where $b_{3}$ is a maximum value $|b'''(\eta)|$ on the segment
$\eta\in (0,2\pi)$. Using (\ref{s:E1}), (\ref{s:Y3}) and obvious
relation $|I_{2}|\leq1$ we find
\begin{eqnarray}
  \label{s:E2}
  \left|\frac{d\dot{E}^{\rm gr}_{l}}{d\Omega}\right|
  \leq8\pi G\mu^{2}\frac{b_{3}^{2}}{l^{2}}.
\end{eqnarray}
Integration of (\ref{s:E2}) over the unit sphere and successive summation
over $l$ using relation $\sum_{l=1}^{\infty}l^{-2}=\pi^{2}/6$ gives the
requested upper limit for the gravitational radiation power
\begin{eqnarray}
  \label{s:E4}
  \left|\dot{E}^{\rm gr}\right| \leq 32\pi^{4} b_{3}^{2}G\mu^{2}.
\end{eqnarray}
Similarly in the case of electromagnetic radiation integrating
(\ref{s:M}) three times by parts and using (\ref{r:Ee2}) we find
\begin{eqnarray}
  \label{s:Ee1}
  \left|\dot{E}^{\rm em}\right|\leq\frac{4}{3}\pi^{4} b_{3}^{2} q^{2}\mu.
\end{eqnarray}
The presence the third derivative $\mathbf{b'''}(\eta)$ in  (\ref{s:E4})
and (\ref{s:Y2}) is not surprise and resemble the quadruple gravitational
radiation formula (see e.~g. \cite{Landau})
\begin{equation}
  \label{s:LL}
  \dot{E}=\frac{G}{45}\dddot{D}_{ij}^{2}
\end{equation}
with the third time derivative of the quadruple momentum $D_{ij}$.
Electromagnetic radiation contains $\ddot{\mathbf{d}}$ in dipole
approach, where $\mathbf{d}$ is the dipole moment. Arguing
similarly it may seem, in this case, that we need restriction on
a second derivative $\mathbf{b}(\eta)$, not the third. But in the
first order of expansion on $k$ dipole radiation is equal to
zero, the first non-zero term is quadruple, therefore we again
obtain dependence from $\mathbf{b}'''$.

Note that requirement of restriction of the third derivative
$\mathbf{b'''}$ is not necessary in general. For example, if the
string has kinks (see below) then the first derivative
$\mathbf{b'}$ is discontinuous (and, consequently
$Y_{p}(l)\propto 1/l$, $M(l)\propto 1/l$). Convergence of series
(\ref{r:sumE}) and (\ref{r:sumEe}) in this case is ensured by the
behavior of fundamental integrals $I_{p}(l)\propto1/l$ at $l\gg1$.

\section{Near stationary loops}
\label{sec:IV}

It is possible to derive rather simple expressions for the total power
radiated by the chiral loops in the limit when loops are very close to
their stationary states, i.~e. $k\ll 1$ in (\ref{in:conab1}).
Additionally it is supposed that $k$ does not depend on $\eta$ and
therefore the current $j^{\mu}$ is constant along the string. Using
expansion of (\ref{r:E3}) and (\ref{r:Ee2}) in powers of $k$, the
corresponding gravitational and electromagnetic power can be written in
the form:
\begin{equation}
  \label{g:E2}
  \dot{E}^{\rm gr}=  K^{\rm gr}G\mu^{2}k^{2}, \quad
  \dot{E}^{\rm em}= K^{\rm em}q^{2}\mu  k^{2},
\end{equation}
where $K^{\rm em}$ and $K^{\rm gr}$ are numerical factors, depending only
on the loop geometry. We see that radiation power of the near stationary
chiral loops is proportional to $k^2$.s The geometrical numerical factor
$K^{\rm gr}$ in turn is connected with the corresponding parameter
$\Gamma$ in equation $\dot E^{\rm gr}=\Gamma G\mu^{2}$ by relation
\begin{equation}
  \label{g:KGamma}
  \Gamma = K^{\rm gr}k^{2}.
\end{equation}

\subsection{Multiply wound chiral loop}
\label{wound}

\begin{figure*}
\includegraphics{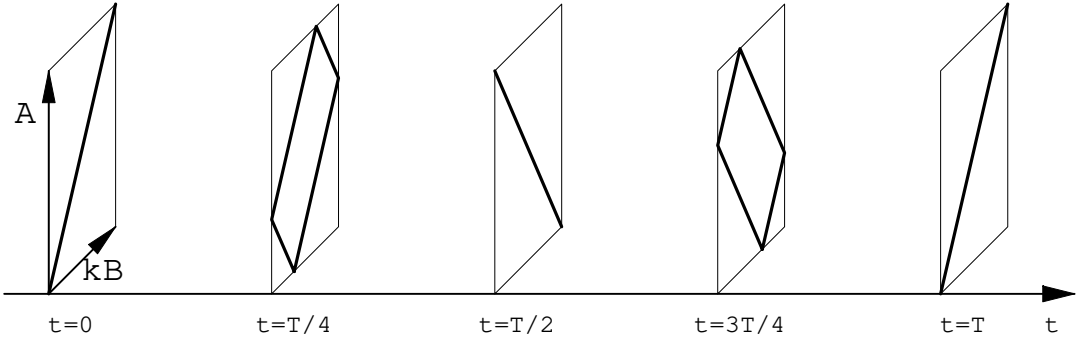}
\caption{\label{fig1} Oscillations of the piece-wise chiral string
(\ref{p:loop1}). When oscillating, string goes in successive sequence
positions with the $T/4$ step, where the oscillation period $T=L/2$.}
\end{figure*}

The first simple solvable example of a near stationary loop radiation is
the multiply wound chiral string \cite{Davis2}:
\begin{eqnarray}
  \label{m:loop1}
  \mathbf{a}&\!=\!&
  \frac{1}{m}\:(\cos m\xi,\; -\sin m\xi, \; 0),\nonumber \\
  \mathbf{b}&\!=\!
  &\frac{k}{n}\:(\cos n\eta, \; -\cos\Phi\sin n\eta, \;
  -\sin\Phi\sin n\eta),
\end{eqnarray}
where $m,n$ are integer (harmonic numbers). For nonzero Fourier component
indexes in (\ref{r:sumE}) and (\ref{r:sumEe}) the following condition is
held: $l=\kappa m=\lambda n$, where $\kappa$ and $\lambda$ are integers
(see Appendix~\ref{app} for details). In all other cases $d\dot{E}^{\rm
gr}_{l}/d\Omega=d\dot{E}^{\rm em}_{l}/d\Omega=0$. Let $\lambda$ be the
minimum integer such as $\kappa m=\lambda n$. In this case first non-zero
term in expanding of the energy on $k$ is proportional to $k^{2\lambda}$:
\begin{eqnarray}
  \label{m:E1}
  \dot E^{\rm gr} & \simeq &
  K^{\rm gr}(\kappa,\lambda,\Phi)l^{2}k^{2\lambda}G\mu^{2}, \nonumber \\
  \dot E^{\rm em} & \simeq &
  K^{\rm em}(\kappa,\lambda,\Phi)l^{2}k^{2\lambda}q^{2}\mu,
\end{eqnarray}
where $K^{\rm gr}(\kappa,\lambda,\Phi)$ and $K^{\rm
em}(\kappa,\lambda,\Phi)$ are numerical factors depending on geometrical
string configuration. For example, the particular case with $m=n=1$ and
$\Phi=0$ in (\ref{m:loop1})  corresponds to pure radial oscillations of
the chiral loop:
\begin{eqnarray}
  \label{c:loop1}
  \mathbf{a}&=& (\cos\xi,\; -\sin\xi,\; 0), \nonumber \\
  \mathbf{b}&=&k(\cos\eta,\; -\sin\eta,\; 0).
\end{eqnarray}
In this case the first non-zero term in expanding of radiated power on
$k$ is proportional to $k^{2}$. Therefore taking into account only first
term in expanding (\ref{r:sumE}) we have (\ref{b:I1})
\begin{eqnarray}
  \label{c:I}
  I_{2}(1)&\!\!=\!\!&\frac{1}{2}\,e^{-i\phi}[J_{2}(-\sin\theta)+\!
  J_{0}(-\sin\theta)]\cos\theta,\nonumber \\
  I_{3}(1)&\!\!=\!\!&\frac{i}{2}\,e^{-i\phi}[J_{2}(-\sin\theta)-\!
  J_{0}(-\sin\theta)].
\end{eqnarray}
By keeping the leading nonzero term at $k\ll 1$ we obtain
(\ref{b:YM2})
\begin{eqnarray}
  \label{c:YM}
  Y_{2}(1)&=&\frac{k}{2}\,e^{i\phi}\cos\theta,\nonumber \\
  Y_{3}(1)&=&i \frac{k}{2}\,e^{i\phi},\nonumber\\
  M(1)&=&\frac{k}{2}\,e^{i\phi}\sin\theta .
\end{eqnarray}
Substituting (\ref{c:I}) and (\ref{c:YM}) in (\ref{r:E3}) and
(\ref{r:Ee2}) we find
\begin{eqnarray}
  \label{c:E1}
  \frac{d\dot{E}^{\rm gr}}{d\Omega}\simeq
  \frac{G\mu^{2}\pi k^{2}}{2}\left\{\left[J_{2}(-\sin\theta)-
  J_{0}(-\sin\theta)\right]^{2}+\right.\nonumber\\
  2\left[3J^{2}_{2}(-\sin\theta)-J^{2}_{0}
  (-\sin\theta)\right]\cos^{2}\theta+ \nonumber \\
  \left.\left[J_{2}(-\sin\theta)+J_{0}(-\sin\theta)\right]^{2}
  \cos^{4}\theta\right\}, \nonumber \\
  \frac{d\dot{E}^{\rm em}}{d\Omega}\simeq
  \frac{\pi k^{2}q^{2}\mu}{8}
  \left\{\left[J_{2}(-\sin\theta)-
  J_{0}(-\sin\theta)\right]^{2}+\right.\nonumber\\
  \left.\left[J_{2}(-\sin\theta)+J_{0}
  (-\sin\theta)\right]^{2}\cos^{2}\theta\right\}\sin^{2}\theta.
\end{eqnarray}
Integration over the unit sphere in (\ref{c:E1}) gives
\begin{eqnarray}
  \label{c:K1}
  K^{\rm gr}=\pi^{2}\int_{0}^{\pi}d\theta \sin \theta
  \left\{\left[J_{2}(\sin\theta)-J_{0}(\sin\theta)\right]^{2}
  +\right.\nonumber\\
  \left. 2\left[3J^{2}_{2}(\sin\theta)
  -J^{2}_{0}(\sin\theta)\right]\cos^{2}\theta+\right. \nonumber \\
  \left.\left[J_{2}(\sin\theta)+J_{0}(\sin\theta)\right]^{2}
  \cos^{4}\theta\right\}, \nonumber \\
  K^{\rm em}=\int_{0}^{\pi}d\theta \sin^{3}\theta
  \left\{\left[J_{2}(\sin\theta)
  -J_{0}(\sin\theta)\right]^{2}+\right.\nonumber\\
  \left.\left[J_{2}(\sin\theta)+J_{0}(\sin\theta)\right]^{2}
  \cos^{2}\theta\right\}.
\end{eqnarray}
with numerical values for $K^{\rm gr}=4.73$ and $K^{\rm em}=2.28$.

\subsection{Piece-wise string evolution}

\begin{figure*}
\includegraphics{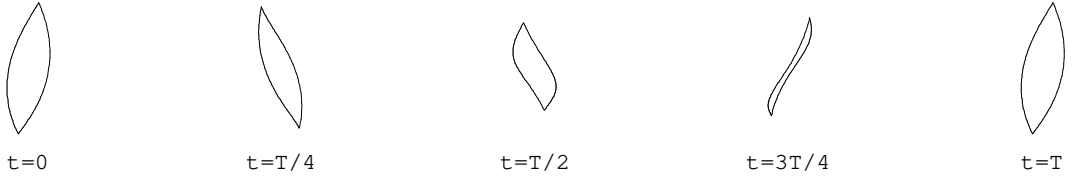}
\caption{\label{fig2} Projections of the stretched string loop
(\ref{d:loop1}) with an invariant length $L$ and oscillation period
$T=L/2$ on the fixed two-dimensional plane. The string is shown in
successive moments of time with the step of $T/4$. The small parameter
$k$ is chosen to be equal $0.5$ for better visuality.
        }
\end{figure*}

Let us consider now the second example of chiral string loop
\begin{eqnarray}
  \label{p:loop1}
  \mathbf{a} &=& \mathbf{A}\left\{
    \begin{array}{lcl}\left(\xi-\pi/2\right), & & 0\le\xi\le\pi, \\
  \left(-\xi+3\pi/2\right), & &   \pi\le\xi\le 2\pi, \\
    \end{array} \right. \nonumber \\
  \mathbf{b} &=& k\mathbf{B}\left\{
    \begin{array}{lcl}\left(\eta-\pi/2\right), & & 0\le\eta\le\pi,\\
  \left(-\eta+3\pi/2\right), & &   \pi\le\eta\le 2\pi, \\
    \end{array} \right.
\end{eqnarray}
where $\mathbf{A}$ and $\mathbf{B}$ are constant unit vectors. This is
the generalization of ordinary piece-wise string loop without the current
\cite{Garfinkle} to the chiral current case. See Fig.~\ref{fig1} for
visualisation of this chiral loop oscillations. For fundamental integrals
(\ref{r:IY1a}) and function $M(l)$ from (\ref{r:M}) we obtain in this
case respectively
\begin{eqnarray}
  \label{p:IYM}
  I_{2}(l)\! &=& \!-\frac{i\mathbf{v}\mathbf{A}
  \left[e^{il\pi k\mathbf{n}\mathbf{B}/2}
  -(-1)^{l}e^{-il\pi k\mathbf{n}\mathbf{B}/2}\right]}
  {\pi l\left[1-(\mathbf{n}\mathbf{A})^{2}\right]},  \\
  I_{3}(l)\! &=& \!-\frac{i\mathbf{w}\mathbf{A}
  \left[e^{il\pi k\mathbf{n}\mathbf{B}/2}
  -(-1)^{l}e^{-il\pi k\mathbf{n}\mathbf{B}/2}\right]}
  {\pi l\left[1-(\mathbf{n}\mathbf{A})^{2}\right]}, \nonumber \\
  Y_{2}(l)&=& \frac{ik\mathbf{v}\mathbf{B}
  \left[e^{il\pi k\mathbf{n}\mathbf{B}/2}
  -(-1)^{l}e^{-il\pi k\mathbf{n}\mathbf{B}/2}\right]}
  {\pi l\left[1-(k\mathbf{n}\mathbf{B})^{2}\right]}, \nonumber \\
  Y_{3}(l)\! &=& \!\frac{ik\mathbf{w}\mathbf{B}
  \left[e^{il\pi k\mathbf{n}\mathbf{B}/2}
  -(-1)^{l}e^{-il\pi k\mathbf{n}\mathbf{B}/2}\right]}
  {\pi l\left[1-(k\mathbf{n}\mathbf{B})^{2}\right]}, \nonumber \\
  M(l)\! &=& \!\frac{ik\mathbf{n}\mathbf{B}(1\!-\! k^{2})^{1/2}\!
  \left[e^{il\pi k\mathbf{n}\mathbf{B}/2}
  -(-1)^{l}e^{-il\pi k\mathbf{n}\mathbf{B}/2}\right]}
  {\pi l\left[1-(k\mathbf{n}\mathbf{B})^{2}\right]}. \nonumber
\end{eqnarray}
Functions $Y_{p}(l)$ decrease here as $1/l$ with growing of $l$.
Combining with similar behavior of $I_{p}(l)\propto 1/l$, this provides
the convergence of energy series in (\ref{r:sumE}) and (\ref{r:sumEe}).
As a result for the gravitational and electromagnetic radiation of the
nearly stationary chiral loop (\ref{p:loop1}) we obtain respectively
\begin{widetext}
\begin{eqnarray}
  \label{p:E1}
  \frac{d\dot{E}^{\rm gr}_{l}}{d\Omega} &=&
  \frac{32G\mu^{2}}{\pi^{3}l^{2}}
  \frac{\left[1-(-1)^{l}\cos(l\pi\mathbf{n}\mathbf{A})\right]
  \left[1-(-1)^{l}\cos(l\pi k\mathbf{n}\mathbf{B})\right]}
  {\left[1-(\mathbf{n}\mathbf{A})^{2}\right]
  \left[1-k^{2}(\mathbf{n}\mathbf{B})^{2}\right]^{2}}
  \left[1-(\mathbf{n}\mathbf{B})^{2}\right]k^{2}, \nonumber \\
  \frac{d\dot{E}^{\rm em}_{l}}{d\Omega} &=&
  \frac{8 q^{2}\mu}{\pi^{3}l^{2}}
  \frac{\left[1-(-1)^{l}\cos(l\pi\mathbf{n}\mathbf{A})\right]
  \left[1-(-1)^{l}\cos(l\pi k\mathbf{n}\mathbf{B})\right]}
  {\left[1-(\mathbf{n}\mathbf{A})^{2}\right]
  \left[1-k^{2}(\mathbf{n}\mathbf{B})^{2}\right]^{2}}
  (\mathbf{n}\mathbf{B})^{2}k^{2}(1-k^{2}),
\end{eqnarray}
\begin{eqnarray}
  \label{p:E1a}
  \frac{d\dot{E}^{\rm gr}}{d\Omega} &=& \frac{16G\mu^{2}}{\pi}
  \frac{1-\frac{1}{2}\left(|\mathbf{n}\mathbf{A}+k\mathbf{n}\mathbf{B}|+
  |\mathbf{n}\mathbf{A}-k\mathbf{n}\mathbf{B}|\right)}
  {\left[1-(\mathbf{n}\mathbf{A})^{2}\right]
  \left[1-k^{2}(\mathbf{n}\mathbf{B})^{2}\right]^{2}}
  \left[1-(\mathbf{n}\mathbf{B})^{2}\right]k^{2}, \nonumber \\
   \frac{d\dot{E}^{\rm em}}{d\Omega} &=&
   \frac{4 q^{2}\mu}{\pi}\frac{1-\frac{1}{2}
   \left(|\mathbf{n}\mathbf{A}+k\mathbf{n}\mathbf{B}|+
  |\mathbf{n}\mathbf{A}-k\mathbf{n}\mathbf{B}|\right)}
  {\left[1-(\mathbf{n}\mathbf{A})^{2}\right]
  \left[1-k^{2}(\mathbf{n}\mathbf{B})^{2}\right]^{2}}
  (\mathbf{n}\mathbf{B})^{2}k^{2}(1-k^{2}).
\end{eqnarray}
\end{widetext}
The last two equations for the radiated energy flux into the solid angle
$d\Omega$ are obtained after summation over $l$ with the using of identity
\cite{Gradshteyn}
\begin{equation}
  \label{p:sumcos}
  \sum_{l=1}^{\infty}\frac{\cos(lx)}{l^{2}}=
  \frac{1}{4}(x-\pi)^{2}-\frac{\pi^{2}}{12}, \quad 0\le x\le 2\pi.
\end{equation}
Note, that (\ref{p:E1}) and (\ref{p:E1a}) are valid for all $k$, not only
for $k\ll 1$. Choosing $A=(1,0,0)$ and $B=(\cos\alpha,\,\sin\alpha,\,0)$
we find in this case from (\ref{g:KGamma}) the corresponding numerical
factors $K^{\rm gr}$ and $K^{\rm em}$:
\begin{eqnarray}
  \label{p:K1}
  K^{\rm gr}& \!\!=\!\! &
  \frac{16}{\pi}\int_{0}^{\pi}\!\!\int_{0}^{2\pi}\!\!
  d\theta d\phi \sin\theta\frac{1\!-
  [\sin\theta\cos(\alpha-\phi)]^2}{1+|\cos\phi\sin\theta|}, \nonumber \\
  K^{\rm em} & \!\!=\!\! &
  \frac{4}{\pi}\int_{0}^{\pi}\!\!\int_{0}^{2\pi}\!\!
  d\theta d\phi \sin\theta
  \frac{[\sin\theta\cos(\alpha-\phi)]^2}{1+|\cos\phi\sin\theta|}.
\end{eqnarray}
For example, in the particular case of $\alpha=\pi/2$ the double
integrals (\ref{p:K1}) equal $K^{\rm gr}=28.3$ and $K^{\rm em}=4$
respectively.

\subsection{Stretched loop oscillation}

Let us consider now the loop of the following kind (combination of the
first and second cases)
\begin{eqnarray}
  \label{d:loop1}
  \mathbf{a} &=& \mathbf{A}\left\{
    \begin{array}{lcl}\left(\xi-\pi/2\right), & &
  0\le\xi\le\pi,\\
  \left(-\xi+3\pi/2\right), & &   \pi\le\xi\le 2\pi, \\
    \end{array} \right. \nonumber \\
  \mathbf{b}&=& k (\sin\eta,\; -\cos\eta,\; 0),
\end{eqnarray}
where $\mathbf{A}$ is the unit vector and $k\ll 1$. We will call this
example as the stretched loop See Fig.~\ref{fig2}. Substituting
(\ref{c:YM}) and (\ref{p:IYM}) in (\ref{r:E3}), and integrating over the
unit sphere, we find
\begin{eqnarray}
  \label{d:K1}
  K^{\rm gr}\!\!\!&=&\!\!\!\frac{12}{\pi}\!\int_{0}^{\pi}\!\!\int_{0}^{2\pi}\!\!\!
  d\theta\:
  d\phi\sin\theta\cos^{2}\theta\frac{[1+\cos\left(\pi\mathbf{n}\mathbf{A}\right)]^{2}}{
  1-(\mathbf{n}\mathbf{A})^{2}},\nonumber \\
  K^{\rm
  em}\!\!\!&=&\!\!\!\frac{1}{2\pi}\!\int_{0}^{\pi}\!\!\int_{0}^{2\pi}\!\!\!
  d\theta\:d\phi\sin^{3}\!\theta \frac{[1+\cos\left(\pi
  \mathbf{n}\mathbf{A}\right)]^{2}}{1-(\mathbf{n}\mathbf{A})^{2}}.
\end{eqnarray}
For example in the case $\mathbf{A}=(0,0,1)$ we find $K^{\rm
gr}=7.63$ and $K^{\rm em}=3$.

\section{Damping of loop oscillations}
\label{sec:V}

Let us evaluate now the damping time of small amplitude oscillations of
near stationary chiral strings corresponding to the limit $k\ll 1$ in
(\ref{in:conab1}). For simplicity we assume that $k$ does not depend on
$\eta$ in the considered limit (this assumption is held true in the
considered above solvable examples). Then a total loop charge
conservation in (\ref{in:Q0}) gives
\begin{equation}
  \label{t:diff0}
  \frac{q\sqrt{\mu}}{2}L\sqrt{1-k^{2}}=const. \nonumber
\end{equation}
From this equation we find the relation between energy $E$ and parameter
$k$ of the chiral string with small amplitude oscillations:
\begin{equation}
  \label{t:diff1}
  E\simeq E_{v}\left(1+\frac{k^{2}}{2}\right),
\end{equation}
where $E_{v}=L\mu$ is the energy of the stationary (vorton) chiral loop
configuration at $k=0$. Comparing (\ref{t:diff1}) with (\ref{g:E2}) we
estimate the damping time of string oscillations
\begin{equation}
  \label{t:tau1}
  \tau\sim\frac{E_{v}}{2(K^{\rm gr}G\mu^{2}+K^{\rm em}q^{2}\mu)}.
\end{equation}
Let us express (\ref{t:tau1}) through vorton length. We have
$E_{v}=L\mu$, where $L$ is the invariant length, and physical length of a
stationary string is equal to half of invariant length $L_{\rm ph}=L/2$
\cite{Steer}. We find
\begin{equation}
  \label{t:tau2}
  \tau\sim\frac{L_{\rm ph}}{K^{\rm gr}G\mu+K^{\rm em}q^{2}}.
\end{equation}
In the case of the chiral ring with one harmonic (\ref{c:loop1}), it is
naturally to assume, that evolving ring saves its form (due to symmetry)
during the damping of small amplitude oscillations. In this case
therefore we can find precisely the evolution of the radiating string
with time. Only $k$ varies in time if the shape of the string is
invariable. Solving together (\ref{g:E2}) and (\ref{t:diff1}), we find
the law of oscillation damping in the near stationary chiral ring
\begin{equation}
  \label{t:sol1}
  k^{2}\simeq k^{2}_{0}e^{-t(1/\tau_{c}^{\rm gr}+1/\tau_{c}^{\rm em})},
\end{equation}
where $k_{0}=k(t=0)$, and correspondingly the damping time due to
gravitational radiation
\begin{eqnarray}
  \label{t:tau3a}
  \tau_{c}^{\rm gr}&=&
  \frac{E_{v}}{2K^{\rm gr}G\mu^{2}}=\frac{L_{\rm ph}}{K^{\rm gr}G\mu}
\end{eqnarray}
and due to electromagnetic radiation
\begin{eqnarray}
  \label{t:tau3b}
  \tau_{c}^{\rm em}&=&
  \frac{E_{v}}{2K^{\rm em}q^{2}\mu}=\frac{L_{\rm ph}}{K^{\rm em}q^{2}}.
\end{eqnarray}
Substituting (\ref{t:sol1}) in (\ref{t:diff1}) we obtain
\begin{equation}
  \label{t:sol2}
  E\simeq E_{v}\left[1+\frac{k^{2}_{0}}{2}
  e^{-t(1/\tau_{c}^{\rm gr}+1/\tau_{c}^{\rm em})}\right].
\end{equation}
An effective number of oscillations during the damping time (oscillator
quality) is
\begin{eqnarray}
  \label{t:Q}
  Q=\frac{\tau}{T}
  =\frac{2}{L}\frac{\tau^{\rm gr}\tau^{\rm em}}
  {\tau^{\rm gr}+\tau^{\rm em}}.
\end{eqnarray}
To restore the usual CGS units we replace $G\mu^{2}\rightarrow
G\mu^{2}c$, $q^{2}\mu\rightarrow q^{2}\mu c^{2}/\hbar$ and choose standard
normalization for the string mass per unit length $G\mu/c^{2}=
10^{-6}\mu_{6}$ and $q_{e}=q/e$ for dimensionless charge carrier on the
string, where elementary electric charge is $e=4.8\times 10^{-10}$. As a
result we obtain for damping time
\begin{eqnarray}
  \label{t:tau4}
  \tau^{\rm gr}=\frac{L_{\rm ph}c}{K^{\rm gr}G\mu}, \quad
  \tau^{\rm em}=\frac{L_{\rm ph}\hbar}{K^{\rm em}q^{2}}.
\end{eqnarray}
The oscillator quality (\ref{t:Q}) in the cases of the gravitational and
electromagnetic radiation are respectively
\begin{eqnarray}
  \label{t:QQ}
   Q^{\rm gr}=
   \frac{1}{K^{\rm gr}}\frac{c^2}{G\mu},\quad
   Q^{\rm em}=
   \frac{1}{K^{\rm em}}\frac{1}{\alpha_{\rm em}q_e^{2}},
\end{eqnarray}
with $\alpha_{\rm em}=e^2/c\hbar$ The ratio of damping times is
\begin{equation}
  \label{t:tt1}
  \tau^{\rm gr}/\tau^{\rm em}\sim
  \frac{q^{2}}{G\mu\hbar}\left(\frac{K^{\rm em}}{K^{\rm gr}}\right)\simeq
  1.4\times10^{-4}\frac{q_{e}^{2}}{\mu_{6}}\frac{K^{\rm em}}{K^{\rm gr}}.
\end{equation}
If $q_{e}^{2}/\mu_{6}\gtrsim 1.4\times 10^{-3}$, then electromagnetic
radiation prevails in the chiral loop evolution (it is valid for the
standard values of $\mu_{6}\sim1$ and $q_{e}\sim1$). If on the contrary
$q_{e}^{2}/\mu_{6}\lesssim 1.4\times 10^{-3}$ (for example, if a current
is neutral and there is no electromagnetic radiation at all), then pure
gravitational radiation determines the evolution.

Let us estimate a characteristic size of the string $L_{\rm v}$ with
oscillation damping time (\ref{t:tau4}) equals to the universe lifetime
$t_{0}\simeq 10^{18}$~s. In the case of the gravitational radiation
predominance we find
\begin{eqnarray}
  \label{t:L1}
  L_{\rm v}^{\rm gr}\simeq \frac{G\mu K^{\rm gr}t_{0}}{c}
  \simeq10^2 \mu_{6} \mbox{ kpc}
\end{eqnarray}
for $K^{\rm gr}\sim1$.  So the chiral strings with length $L<L_{\rm
v}^{\rm gr}$ (i.~e. with a size of typical galactic halo or less) have
had enough time to fade into vortons. On the other hand if the
electromagnetic radiation prevails, we have
\begin{equation}
  \label{t:Le1}
  L_{\rm v}^{\rm em}\simeq \frac{q^{2} K^{\rm em}t_{0}}{\hbar}
  \simeq70 q_{e}^{2} \mbox{ Mpc}
\end{equation}
for $K^{\rm em}\sim1$ and so the electromagnetically radiated chiral
loops with the length shorter than the size of galactic clusters have had
transformed now to vortons.

\section{Discussion}
\label{sec:VI}

We have explored the asymptotic fading of chiral cosmic string loops into
vortons. It turns out that the upper bound of the gravitational radiation
of a near stationary loop (\ref{s:Ee1}) is proportional to the squared
third derivative of the oscillation amplitude. This result is in
qualitative agreement with the known formula for gravitational radiation
in quadruple approximation.

The derived equations (\ref{r:E3}) and (\ref{r:Ee2}) for respectively the
gravitational and electromagnetic power radiation of oscillating chiral
loops are suitable for practical calculations. We demonstrate that, if
oscillations are small in amplitude ($k\ll 1 $) and the current $j^{\mu}$
is constant along the string, then the power of the gravitational and
electromagnetic radiation (\ref{g:E2}) is proportional to $k^{2}$ and the
coefficient of proportionality depends only on the form of the loop. For
some examples of chiral loops we calculated the total radiated power in
the limit of small amplitude oscillations. For chiral ring with small
amplitude radial oscillations the radiated power per solid angle
$d\Omega$ for the electromagnetic (\ref{p:E1}) and gravitational
(\ref{p:E1a}) radiation is found analytically.

We estimated the damping time of chiral loops (\ref{t:tau1}) with small
amplitude oscillations. In the case of the gravitational radiation
prevalence this time is $\tau^{\rm gr}\sim L_{\rm ph}/K^{\rm gr} G\mu $,
where $K^{\rm gr}$ is a numerical coefficient depending on the string
geometry. The damping time due to the gravitational radiation of the
considered chiral loops is of the order of magnitude longer than the
lifetime of ordinary cosmic strings. If the electromagnetic radiation is
prevail then decay time is $\tau^{\rm em}\sim L_{\rm ph}/K^{\rm em}
q^{2}$. The oscillation energy and amplitude parameter $k$ of the radially
oscillating chiral ring are found to be exponentially decaying with time
according to respectively (\ref{t:sol1}) and (\ref{t:sol2}).

From the damping time estimation it follows that only sufficiently long
superconducting cosmic strings oscillate up to the present time. On the
contrary the small scale chiral loops transformed into the stationary
vortons due to the oscillation damping. Namely, the minimal length of
presently oscillating chiral loop varies from $L_{\rm v}^{\rm gr}\sim
10^2 \mu_{6}$~kpc for gravitational radiation domination to $L_{\rm
v}^{\rm em}\sim70q_{e}^{2}$~Mpc for electromagnetic radiation domination
depending on the relations between $\mu$ and $q$. It appears that the
oscillator quality of chiral loops (\ref{t:QQ}) is independent on the
loop length and is determined only by the loop shape through geometric
parameters $K^{\rm gr}$ and $K^{\rm em}$. For characteristic values of
$K^{\rm gr}\sim 10$ and $K^{\rm em}\sim 1$ the corresponding oscillator
qualities for the gravitational and electromagnetic radiation are $Q^{\rm
gr}\sim 10^{5}/\mu_{6}$ and $Q^{\rm em}\sim 137/q_{e}^{2}$.

\begin{acknowledgments}
This work was supported in part by Russian Foundation for basic Research
grants 00-15-96632 and 00-15-96697 and by the INTAS through grant 99-1065.
\end{acknowledgments}

\appendix
\section{Multiply wound ring}
\label{app}

Here we calculate the fundamental integrals (\ref{r:I1}), (\ref{r:Y1})
and (\ref{r:M}) which define the radiated power of the string for the
particular case of multiply wound loop in the Section~\ref{wound}. We
begin from the consideration of the following integrals
\begin{eqnarray}
  \label{a:1}
  Is_{lm}(x,y)\!\!&=\!\!&\frac{1}{2\pi}\!\int_{0}^{2\pi}\!\!\!\!d\eta
  e^{il(\eta+x\sin m\eta+y\cos m\eta)}\sin\eta,\nonumber\\
  Ic_{lm}(x,y)\!\!&=&\!\!\frac{1}{2\pi}\!\int_{0}^{2\pi}\!\!\!\!d\eta
  e^{il(\eta+x\sin m\eta+y\cos m\eta)}\cos\eta.\quad
\end{eqnarray}
The combinations $Is_{lm}\pm Ic_{lm}$ are integrals of the form
\begin{equation}
  \label{a:2}
  S_{nm}(x,y)=\frac{1}{2\pi}\int_{0}^{2\pi} \!\!d\eta
  e^{il(\eta+x\sin m\eta+y\cos m\eta)},
\end{equation}
and integral (\ref{a:2}) is expressed as follows
\begin{equation}
  \label{a:3}
  S_{nm}(x,y)\! =\! \left\{\begin{array}{ll}
  \! \left(\frac{x-i y}{r}\right)^{n/m} T_{nm}(n r),& r\neq 0, \\
  \! 0,&  r=0, n\neq 0, \\
  \! 1,&  r=0, n=0, \end{array}\right.
\end{equation}
where $r=(x^{2}+y^{2})^{1/2}$ and
\begin{equation}
  \label{a:4}
  T_{nm}(r)=\frac{1}{2\pi}\int_{0}^{2\pi} d\eta
  e^{i(n\eta+r\sin m\eta)}.
\end{equation}
Let $n,m$ are integer. If $n=c m$, where $c$ is integer, then
replacing $\eta\rightarrow m\eta $ and taking into account, that
the function in integral (\ref{a:4}) is periodic we find
\begin{equation}
  \label{a:5}
  T_{nm}(r)=J_{c}(-r),\quad n=cm, \quad n,m,c\in Z,
\end{equation}
where $J_{c}$ is $c$-th Bessel function. Then integrals (\ref{a:1}) can be
expressed through Bessel functions in the following way
\begin{eqnarray}
  \label{a:6}
  Is_{nm}\!\!&=\!\!&-\frac{i}{2}\!
  \left(\frac{z^*}{r}\right)^{c}
  \left[\frac{z^*}{r}J_{c+1}(-cr)-\!
  \frac{z}{r}J_{c-1}(-cr)\right]\!, \nonumber\\
  Ic_{nm}\!\!&=\!\!&\frac{1}{2}\! \left(\frac{z^*}{r}\right)^{c}
  \left[\frac{z^*}{r}J_{c+1}(-cr)+\!
  \frac{z}{r}J_{c-1}(-cr)\right]\!, \quad
\end{eqnarray}
where $z=x+i y$ and $z^*=x-i y$. Let us demonstrate now that if $n\neq c
m$, where $c$ is integer, then integrals (\ref{a:1}) equal zero. In other
words we need to prove, that $T_{nm}(r)\equiv 0$ if $n\neq c m$. First of
all we note that
\begin{eqnarray}
  \label{a:7}
  f(\eta)={\rm Im}\, e^{i(n\eta+r\sin m\eta)}=-f(2\pi-\eta)
\end{eqnarray}
and therefore ${\rm Im}\, T_{nm}(r)\equiv 0$. The real part of function in
the integral (\ref{a:4}) is
\begin{equation}
  \label{a:8}
  \cos n\eta\cos(r \sin m\eta)-\sin n\eta\sin(r \sin m\eta).
\end{equation}
Expansion of $\cos(r \sin m\eta)$ and $\sin(r \sin m\eta)$ in Taylor
series gives:
\begin{eqnarray}
  \label{a:9}
  \cos(r\sin m\eta)&=&
  \sum_{k=0}^{\infty}(-1)^{k}\frac{(r\sin m\eta)^{2 k}}{2k!}, \nonumber \\
  \sin(r\sin m\eta)&=&\sum_{k=0}^{\infty}(-1)^{k}\frac{(r\sin
  m\eta)^{2 k+1}}{(2k+1)!},
\end{eqnarray}
where for odd and even values of $k$ we have the standard binomial-like
representation
\begin{widetext}
\begin{eqnarray}
  \sin^{k}m\eta =\!
  \left(\frac{1}{2i}\right)^{k-1}\left[\sin km\eta
  \!-\!C_{1}^{k}\sin(k\!-\!2)m\eta\!+...
  +\!(-1)^{\frac{k-1}{2}}C_{(k-1)/2}^{k}\sin m\eta\right], \nonumber \\
  \sin^{k}m\eta =\!
  \frac{(-1)^{\frac{k}{2}}}{2^{k-1}} \left[\cos
  km\eta\!+...
  +\!(-1)^{\frac{k-1}{2}}C_{(k-1)/2}^{k}\cos
  2 m\eta\right]\!+\!\frac{1}{2^{k}}C_{k/2}^{k}. \,
\end{eqnarray}
\end{widetext}
It can be seen that only terms of the kind
\begin{eqnarray}
  \int_{0}^{2\pi}\!\!\! d\eta\,\sin (n\eta)\, \sin[(2k+1)m\eta],
  \nonumber\\
  \int_{0}^{2\pi}\!\!\! d\eta\,\cos (n\eta)\, \cos(2k m\eta),
\end{eqnarray}
enter in the considered series (\ref{a:9}). Using this fact and
(\ref{a:7}) we find, that if $n\neq c m$, where $c$ is integer, then
integrals (\ref{a:1}) equal zero, what we wanted to prove.

Let us consider the following chiral circular loop:
\begin{eqnarray}
  \label{b:loop1}
  \mathbf{a}&\!=\!& \frac{1}{m}(\cos m\xi-\sin m\xi),\nonumber \\
  \mathbf{b}&\!\!\!=\!\!\!&\frac{k}{n}(\cos n\eta-\cos\Phi\sin n\eta
  -\sin\Phi\sin n\eta), \quad
\end{eqnarray}
where $m,n$ are integer. For $l=\kappa m$, where $\kappa$ is
integer, we obtain (see (\ref{a:1},\ref{a:6}))
\begin{eqnarray}
  \label{b:I1}
  I_{2}(l)&\!\!=\!\!&\frac{1}{2}\cos\theta[J_{\kappa+1}
  (-\kappa\sin\theta)\!+\!
  J_{\kappa-1}(-\kappa\sin\theta)]e^{-i\kappa\phi},\nonumber \\
  I_{3}(l)&\!\!=\!\!&
  \frac{i}{2}[J_{\kappa+1}(-\kappa\sin\theta)\!-\!
  J_{\kappa-1}(-\kappa\sin\theta)]e^{-i\kappa\phi}.
\end{eqnarray}
For all others $l$ we have $I_{p}(l)=0$. Denoting further
$x=k\:\sin\theta\:\cos\phi$, $y=k(\cos\Phi\:\sin\theta\:\sin\phi +
\sin\Phi\:\cos\theta)$, $r=(x^{2}+y^{2})^{1/2}$, we can write
\begin{eqnarray}
  \label{b:YM1}
  Y_{2}(l)&\!\!=\!\!&k\cos\theta\cos\phi Ic_{l,n}(-x,y)\nonumber \\
  &&+k(\cos\Phi\cos\theta\sin\phi-\sin\Phi\sin\theta)
  Is_{l,n}(-x,y),\nonumber\\
  Y_{3}(l)&\!\!=\!\!&-k\sin\phi Ic_{l,n}(-x,y)+k\cos\Phi\cos\phi
  Is_{l,n}(-x,y),\nonumber\\
  M(l)&\!\!=\!\!&S_{l,n}(-x,y)\sqrt{1-k^{2}}.
\end{eqnarray}
If $l=\lambda n$, where $\lambda$ is integer, then $Y_{p}\neq 0$,
$M(l)\neq 0$, In all other cases $Y_{p}=M(l)= 0$. As a result: if
$l=\kappa m=\lambda n$ (where $\kappa$, $\lambda$ are integer), then
$l$-th term in expanding of the energy is nonzero. In all other cases
$d\dot{E}^{\rm gr}_{l}/d\Omega=d\dot{E}^{\rm em}_{l}/d\Omega=0$. Let
$\lambda$ is minimal integer, such as $\kappa m=\lambda n$. From the
properties of Bessel functions
\begin{equation}
  \label{b:bessel1}
  J_{l}(x)=\frac{x^{l}}{2^{l}l!}+O(x^{2}),
\end{equation}
it follows for (\ref{a:3}) and (\ref{a:6}) in the first nonzero term of
expansion on $k$:
\begin{eqnarray}
  \label{b:Isc}
  Is_{ln}&=&\frac{i}{2}(-x-iy)^{\lambda-1}
  \frac{(-\lambda)^{\lambda-1}}{2^{\lambda-1}(\lambda-1)!},\nonumber\\
  Ic_{ln}&=&\frac{1}{2}(-x-iy)^{\lambda-1}
  \frac{(-\lambda)^{\lambda-1}}{2^{\lambda-1}(\lambda-1)!},\nonumber\\
  S_{ln}&=&(-x-iy)^{\lambda}
  \frac{(-\lambda)^{\lambda}}{2^{\lambda}(\lambda)!}.
\end{eqnarray}
Using (\ref{b:Isc}), we obtain (\ref{b:YM1})
\begin{eqnarray}
  \label{b:YM2}
  Y_{2}(l)&\!=\!&\frac{k}{2}(-x-iy)^{\lambda-1}
  \frac{(-\lambda)^{\lambda-1}}{2^{\lambda-1}(\lambda-1)!}
  \times\nonumber\\
  &&\left(\cos\theta\cos\phi+i\cos\Phi\cos\theta\sin\phi-
  i\sin\Phi\sin\theta\right),\nonumber\\
  Y_{3}(l)&\!=\!&\frac{k}{2}(-x-iy)^{\lambda-1}
  \frac{(-\lambda)^{\lambda-1}}{2^{\lambda-1}(\lambda-1)!}\times\nonumber\\
  &&\left(-\sin\phi +i\cos\Phi\cos\phi \right),\nonumber\\
  M(l)&=&(-x-iy)^{\lambda}
  \frac{(-\lambda)^{\lambda}}{2^{\lambda}(\lambda)!}\sqrt{1-k^{2}}.
\end{eqnarray}
By substituting (\ref{b:YM2}) in (\ref{r:E3}) and (\ref{r:Ee2})
we can calculate the corresponding parameter $\Gamma$ from
expressions for $\dot{E}^{\rm gr}=\Gamma^{\rm gr} G\mu^{2}$ and
$\dot{E}^{\rm em} = \Gamma^{\rm em} G\mu^{2}$.


\begin{thebibliography}{9}

\bibitem{Kibble1} T. W. B. Kibble, J. Phys. {\bf A 9}, 1387 (1976);
    T. W. B. Kibble, G. Lazarides and Q. Shafi, Phys. Rev. {\bf D 26}, 435 (1982)

\bibitem{Zel'dovich1} Y. B. Zel'dovich, Mon. Not. R. Astron. Soc.
    {\bf 192}, 663 (1980)

\bibitem{Vilenkin1} A. Vilenkin, Phys. Rev. D 24, 2082 (1981);
    A. Vilenkin, Phys. Rep. {\bf 121}, 263 (1985)

\bibitem{Witten1} E. Witten, Nucl. Phys. {\bf B249}, 557 (1985)

\bibitem{Davis1} R. L. Davis and E. P. S. Shellard, Phys. Lett.
    {\bf B209}, 485 (1988)

\bibitem{Haws1} D. Haws, M. Hindmarsh and N. Turok, Phys. Lett.
   {\bf B209}, 255 (1988)

\bibitem{Haws2} E. Copeland, D. Haws, M. Hindmarsh and N. Turok,
    Nucl. Phys. {\bf B306}, 908 (1988)

\bibitem{Vilenkin2} E. P. S. Shellard and A. Vilenkin, Cosmic Strings
    and other Topological Defects (Cambridge University Press, Cambridge,
    England, 1994)

\bibitem{Kibble2} M. B. Hindmarsh and T. W. B. Kibble, Rep. Prog. Phys.
   {\bf 58}, 477 (1995)

\bibitem{Carter1} B. Carter and P. Peter, Phys. Lett. {\bf B466}, 41 (1999)

\bibitem{Davis2} A. C. Davis, T. W. B. Kibble, M. Pickles and D.A. Steer,
    Phys. Rev. {\bf D 62} (2000) 083516

\bibitem{Vilenkin3} J.J. Blanco-Pillado, Ken D. Olum and A. Vilenkin,
    Phys. Rev. {\bf D 63}, 103513

\bibitem{Blanco} J.J. Blanco-Pillado and Ken D. Olum, Nucl. Phys.
   {\bf B599} (2001) 435

\bibitem{Berezinsky}  V. Berezinsky, B. Hnatyk and A. Vilenkin,
    Phys. Rev. {\bf D 64} (2001) 043004

\bibitem{Quashnock} J.M. Quashnock and D.N. Spergel, Phys. Rev.
   {\bf D 42} (1990) 2505

\bibitem{Weinberg} S. Weinberg, Gravitation and Cosmology
    (Wiley, New York, 1974)

\bibitem{Durrer} R. Durrer, Nucl. Phys. {\bf B328}, 238 (1989)

\bibitem{Garfinkle} D. Garfinkle and T Vachaspati, Phys. Rev.
   {\bf D 36}, 2229 (1987)

\bibitem{Gradshteyn} I.S. Gradshteyn and I.M. Ryzhik, Tables of Integrals,
    Series and Products (Academic, New York, 1980)

\bibitem{Steer} D.A. Steer, Phys.Rev. {\bf D 63}, (2001) 083517

\bibitem{Landau} L.D. Landau and E.M. Lifshitz, The Classical Theory of Fields.

\bibitem{Vilenkin4} A. Vilenkin and T. Vachaspati, Phys.Rev.Lett
{\bf 58}, 1041 (1987)

\end{thebibliography}
\end{document}